\documentclass{article}
\usepackage{amssymb}

\usepackage{cite}
\usepackage{graphicx}
\usepackage{dcolumn}

\begin{document}

\date{}
\title{On the application of statistical mechanics to diatomic molecules}
\author{Francisco M. Fern\'{a}ndez\thanks{%
fernande@quimica.unlp.edu.ar} \\
INIFTA, DQT, Sucursal 4, C.C 16, \\
1900 La Plata, Argentina}
\maketitle

\begin{abstract}
We argue that the results proposed recently for the $\mathrm{a}^3\Sigma_u^+$
electronic state of $^7\mathrm{Li}_2$ do not exhibit any physical utility
because the canonical vibrational partition function for an excited
electronic state makes no sense according to the principles of statistical
mechanics. Such an approach may, in principle, be applied to the ground
state of a diatomic molecule. We show that the simpler model based on the
harmonic oscillator and rigid rotor enables one to derive a theoretical
expression for the equilibrium constant of a simple chemical reaction that
agrees remarkably well with experimental data. It is not clear whether the
complicated formulas mentioned above are suitable for any comparison with
experimental data.
\end{abstract}

\section{Introduction}

\label{sec:intro}

In a recent paper Jia et al\cite{JZW17} derived analytical expressions for
the vibrational mean energy, vibrational specific heat, vibrational free
energy and vibrational entropy for diatomic molecules. To this end they
resorted to the analytic vibrational energy eigenvalues obtained from the
Schr\"{o}dinger equation with the so-called improved Manning-Rosen
potential-energy function. In order to obtain analytical expressions for
those thermodynamic functions the authors resorted to the Poisson summation
formula\cite{S07}. As a particular example Jia et al\cite{JZW17} chose the
state $\mathrm{a}^{3}\Sigma _{u}^{+}$ of $^{7}\mathrm{Li}_{2}$.

The purpose of this paper is the analysis of the validity of such results.
In section~\ref{sec:CPF} we briefly address the problem of deriving the
canonical partition function for a diatomic molecule. In section~\ref{sec:KP}
we summarize the textbook calculation of the equilibrium constant for a
simple chemical reaction and compare theoretical results with available
experimental data. Finally, in section~\ref{sec:conclusions} we summarize
the main results and draw conclusions.

\section{The canonical partition function for a diatomic molecule}

\label{sec:CPF}

By means of the well known Born-Oppenheimer approximation one commonly
solves the Schr\"{o}dinger equation for a diatomic molecule in two steps\cite
{P68}. In the first one, one solves an eigenvalue equation for the electrons
for different nuclear configurations and obtains the electronic
potential-energy curves $U_{k}(r)$, $k=0,1,\ldots $. In the second step, one
solves a Schr\"{o}dinger equation for the nuclei in every potential-energy
curve; for example
\begin{equation}
H_{k}\psi _{k,J,\nu }=E_{k,J,\nu }\psi _{k,J,\nu },\;H_{k}=-\frac{\hbar ^{2}%
}{2\mu }\nabla ^{2}+U_{k}(r),  \label{eq:Schro}
\end{equation}
where $\mu $ is the nuclear reduced mass and $J=0,1,\ldots $ and $\nu
=0,1,\ldots $ are the rotational and vibrational quantum numbers,
respectively. In order to simplify the discussion we suppose that all the
electronic potential-energy curves required for the calculation support
bound states. Besides, we do not consider the nuclear statistics in order to
facilitate the comparison with the results of Jia et al\cite{JZW17}.

According to the well known principles of statistical mechanics, the
canonical partition function is given by\cite{M76}
\begin{equation}
Q=\sum_{k}g_{e,k}\sum_{J_{k}}\sum_{\nu _{k}}\left( 2J_{k}+1\right) \exp
\left( -\beta E_{k,J,\nu }\right) ,  \label{eq:Q_general}
\end{equation}
where $g_{e,k}$ is the degeneracy of the $k$-th electronic state, $\beta
=1/(k_{B}T)$, $k_{B}$ is the Boltzmann constant and $T$ the absolute
temperature. All the thermodynamic functions can be obtained from $\ln Q$%
\cite{M76}.

At sufficiently low temperatures $\exp \left( -\beta E_{k,J,\nu }\right) \ll
\exp \left( -\beta E_{0,J,\nu }\right) $, $k=1,2,\ldots $ and only the
ground electronic state contributes:
\begin{equation}
Q\approx Q_{0}=g_{e,0}\sum_{J}\sum_{\nu }\left( 2J+1\right) \exp \left(
-\beta E_{0,J,\nu }\right) .  \label{eq:Q_0}
\end{equation}
At higher temperatures, we should add excited electronic states but the
ground state is always present. Therefore, the calculation of thermodynamic
functions with the canonical vibrational partition function based only on an
excited electronic state makes no sense\cite{M76}. In the case of $^{7}%
\mathrm{Li}_{2}$ the $\mathrm{a}^{3}\Sigma _{u}^{+}$ electronic state is an
excited one and, consequently, the results derived by Jia et al\cite{JZW17}
do not have any physical utility. In principle, it would have been
reasonable to carry out the calculation with the experimental data for the
ground electronic state $X^{1}\Sigma _{g}^{+}$\cite{H50}.

Since the approach proposed by Jian et al\cite{JZW17} may be of potential
utility when properly applied to the ground electronic state of a diatomic
molecule, in what follows we briefly summarize the equations that will arise
in such a situation. For concreteness we assume that we have the energy
eigenvalues $0<E_{0}<E_{1}<\cdots <E_{M}$ and define the sums
\begin{equation}
S_{m}(\beta )=\sum_{n=0}^{M}E_{n}^{m}\exp \left( -\beta E_{n}\right)
,\;m=0,1,\ldots .  \label{eq:S_m}
\end{equation}
All these sums are positive, monotonically increasing functions of $T$
\begin{equation}
\frac{\partial }{\partial T}S_{m}=k_{B}\beta ^{2}S_{m+1},
\end{equation}
that satisfy $\lim\limits_{T\rightarrow 0}S_{m}=0$ and $\lim\limits_{T%
\rightarrow \infty }S_{m}=S_{m}(0)$. Obviously, the canonical partition
function and the mean energy can be expressed in terms of these sums as $%
S_{0}=Q$ and $U=S_{1}/S_{0}$, respectively. It is clear that one should add
the continuum spectrum to the partition function but for simplicity we
assume that this contribution is negligible.

A straightforward calculation shows that the specific heat is given by
\begin{equation}
C=\frac{\partial U}{\partial T}=k_{B}\beta ^{2}\left( \frac{S_{2}}{S_{0}}-%
\frac{S_{1}^{2}}{S_{0}^{2}}\right) =\frac{k\beta ^{2}}{S_{0}}%
\sum_{n=0}^{M}\left( E_{n}-U\right) ^{2}\exp \left( -\beta E_{n}\right) >0.
\label{eq:C=dU/dT}
\end{equation}
We appreciate that $U$ increases monotonously from $E_{0}$ to $%
S_{1}(0)/(M+1) $. When $T\rightarrow 0$ the specific heat tends to zero
exponentially as $k_{B}\beta ^{2}\Delta ^{2}e^{-\beta \Delta }$, where $%
\Delta =E_{1}-E_{0}$. On the other hand, when $T\rightarrow \infty $ the
specific heat tends to zero as $T^{-2}$. Therefore, $C$ should exhibit a
maximum somewhere between $0<T<\infty $. This simple mathematical analysis
accounts for some of the numerical results obtained by Jia et al\cite{JZW17}.

The following question arises: what expressions are simpler, those given in
terms of the sums $S_{m}$ or the analytical ones provided by Jia et al?. In
the former case we can directly resort to the experimental data given by the
vibrational-rotational spectrum of the diatomic molecule. In the latter
case, one should derive the parameters of the improved Manning-Rosen
potential from experimental data, solve the Schr\"{o}dinger equation, and
finally obtain the thermodynamic functions either from the sums shown above
or from the cumbersome expressions derived from the Poisson summation
formula and the expansion for $\beta \ll 1$\cite{JZW17}. Of course,
adjectives like simple, straightforward, practical, etc. are a matter of
taste.

Jia et al did not compare their theoretical results with any experimental
data because their expressions do not exhibit any physical utility as argued
above. In the next section we show an application of statistical mechanics
to an actual chemical problem.

\section{Equilibrium constant for a simple chemical reaction}

\label{sec:KP}

For concreteness, we choose the chemical reaction
\begin{equation}
2\mathrm{Na}\rightleftarrows \mathrm{Na}_{2}.  \label{eq:reaction}
\end{equation}
The equilibrium constant is given by\cite{M76}
\begin{equation}
K_{P}(T)=\frac{p_{\mathrm{Na}_{2}}}{p_{\mathrm{Na}}^{2}}=\left(
k_{B}T\right) ^{-1}\frac{q_{\mathrm{Na}_{2}}/V}{\left( q_{\mathrm{Na}%
}/V\right) ^{2}},  \label{eq:K_P}
\end{equation}
where
\begin{eqnarray}
\frac{q_{\mathrm{Na}}\left( T,V\right) }{V} &=&\left( \frac{2\pi m_{\mathrm{%
Na}}k_{B}T}{h^{2}}\right) ^{3/2}g_{e}\left( \mathrm{Na}\right)  \nonumber \\
\frac{q_{\mathrm{Na}_{2}}\left( T,V\right) }{V} &=&\left( \frac{2\pi m_{%
\mathrm{Na}_{2}}k_{B}T}{h^{2}}\right) ^{3/2}\left( \frac{T}{2\theta _{r}}%
\right) \left( 1-e^{-\theta _{v}/T}\right) ^{-1}  \nonumber \\
&&\times g_{e}\left( \mathrm{Na}_{2}\right) e^{D_{0}/\left( k_{B}T\right) }
\label{eq:q's}
\end{eqnarray}
In this expression $m_{A}$ and $g_{e}(A)$ are the mass and electronic
degeneracy, respectively, of the species $A$, $V$ the volume, $\theta _{r}$
and $\theta _{v}$ the characteristic temperatures of rotation and vibration,
respectively, and $D_{0}$ the dissociation energy\cite{M76}.

On the other hand, Ewing et al\cite{ESSM67} fitted experimental data to the
following empirical expression
\begin{equation}
\log K_{P}=-4.3249+\frac{4002.3}{T}.  \label{eq:logK_P_exp}
\end{equation}

\begin{figure}[tbp]
\begin{center}
\includegraphics[width=9cm]{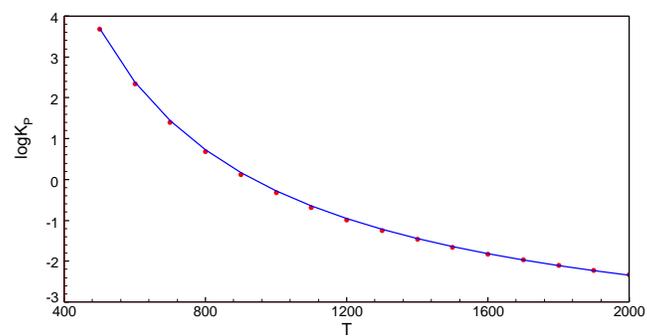}
\end{center}
\caption{Experimental data (red circles) and theoretical results (blue line)
for the equilibrium constant for the chemical reaction (\ref{eq:reaction}) }
\label{Fig:logKP}
\end{figure}

Figure~\ref{Fig:logKP} shows that the theoretical results given by the
textbook formula (\ref{eq:K_P}) (blue line) agree with the empirical formula
(\ref{eq:logK_P_exp}) (red circles) in the whole range of temperature values
considered. According to Ewing et al\cite{ESSM67} the empirical formula (\ref
{eq:logK_P_exp}) was derived from experimental data in the temperature range
$1135\leq t\left( ^{0}C\right) \leq 1413$. However, figure~\ref{Fig:logKP}
suggests that its range of validity is considerably larger.

The analysis just given clearly shows the utility of the well
known textbook model based on the rigid rotor and harmonic
oscillator. The remarkable agreement of the theoretical results
thus obtained with available experimental data enabled us to argue
that the range of validity of the empirical formula
(\ref{eq:logK_P_exp}) is greater than expected. It is not clear to
us if the complicated formulas derived by Jia et al\cite{JZW17}
from a different model are suitable for a successful comparison
with experimental data.

\section{Conclusions}

\label{sec:conclusions}

The results derived by Jia et al\cite{JZW17} for the $a^{3}\Sigma _{u}^{+}$
of $^{7}\mathrm{Li}_{2}$ are of no physical utility because the vibrational
energies of an excited electronic state do not appear alone in the
calculation of the thermodynamic functions of diatomic molecules. We cannot
calculate thermodynamic functions for particular electronic states, except
for the ground one at sufficient low temperatures. The results of Jia et al
may be useful for the study of the thermodynamic functions of diatomic
molecules at sufficiently low temperatures if we apply the approach to the
ground electronic state. Even in such a case the question arises as to the
practicality of such expressions compared with well known alternative ones
as argued above.

The standard model given by the rigid rotor and harmonic oscillator commonly
suffices for many physical applications as shown by the results in section~%
\ref{sec:KP}. The agreement between such theoretical results and available
experimental data is remarkable and enabled us to test the empirical formula
(\ref{eq:logK_P_exp}) beyond the temperature range used to derive it. It is
not clear to us whether the theoretical results proposed by Jia et al\cite
{JZW17} are suitable for any comparison with experimental data.

\end{document}